\documentclass[prb,twocolumn,showpacs,preprintnumbers,superscriptaddress,amsmath,amssymb]{revtex4-1}
\usepackage{graphicx}
\usepackage{verbatim}

\begin{document}

%\title{Emergence of spin-polaron peaks in optical spectra of antiferromagnet: a fingerprint of strong electronic correlations}
%\title{Can optical spectroscopy detect the emergence of antiferromagnetic order\\ 
%in strongly correlated electron systems?}
%AG If a queston is not acceptable in the title by PRB:
\title{Signature of antiferromagnetic long-range order in the optical spectrum\\ 
of strongly correlated electron systems}

\author{C.~Taranto}
\affiliation{Institute for Solid State Physics, Vienna University of Technology, 1040 Vienna, Austria}
\author{G.~Sangiovanni}
\affiliation{Institute for Solid State Physics, Vienna University of Technology, 1040 Vienna, Austria}
\author{K.~Held}
\affiliation{Institute for Solid State Physics, Vienna University of Technology, 1040 Vienna, Austria}
\author{M.~Capone}
\affiliation{Democritos National Simulation Center, Consiglio Nazionale delle Ricerche, Istituto Officina dei Materiali (IOM) and Scuola Internazionale Superiore di Studi Avanzati (SISSA), Via Bonomea 265, 34136 Trieste, Italy}
\author{A.~Georges}
\affiliation{Centre de Physique Th\'eorique, Ecole Polytechnique, CNRS, 91128 Palaiseau Cedex, France}
\affiliation{Coll\`ege de France, 11 place Marcelin Berthelot, 75005 Paris, France}
\affiliation{DPMC, Universit\'e de Gen\`eve, 24 quai Ernest Ansermet, CH-1211 Gen\`eve, Suisse}
\author{A.~Toschi}
\affiliation{Institute for Solid State Physics, Vienna University of Technology, 1040 Vienna, Austria}

\date{Version 1, \today }

\begin{abstract}
We show how the onset of a non-Slater antiferromagnetic ordering in a correlated material can be detected by optical spectroscopy. Using dynamical mean-field theory we identify two distinctive features: The antiferromagnetic ordering is associated with an enhanced spectral weight above the optical gap, and well separated spin-polaron peaks emerge in the optical spectrum.
Both features are indeed observed in LaSrMnO$_4$ [G\"ossling {\it et al.}, Phys. Rev. B {\bf 77}, 035109 (2008)].
\end{abstract}

\pacs{71.27.+a, 71.10.Fd}

\maketitle

\let\n=\nu \let\o =\omega \let\s=\sigma

% 71.27.+a  Strongly correlated electron systems; heavy fermions
% 71.10.Fd  Lattice fermion models (Hubbard model, etc.)
% 71.30.+h  Metal-insulator transitions and other electronic transitions
% 71.20.Eh  Rare earth metals and alloys
% 75.20.Hr  Local moment in compounds and alloys; Kondo effect, valence
%           fluctuations, heavy fermions
% 75.47.Gk  Colossal magnetoresistance

\section{Introduction}
This article deals with the following issue: Is there a distinctive signature of the onset of antiferromagnetic (AF) long-range order in the optical spectrum of a strongly correlated insulating antiferromagnet, for example an antiferromagnetic transition-metal oxide? 

AF ordering is a characteristic feature of strongly correlated Mott insulators where superexchange drives the ordering of localized magnetic moments. On the other hand, AF can also arise from Fermi-surface nesting in a weakly correlated material. From a theoretical point of view, a continuous crossover driven by the interaction strength connects weak- and strong-coupling antiferromagnets, and an unambiguous distinction between local-moments and Fermi-surface AF is lacking.

In this work we propose that, remarkably, optical spectroscopy can be used to infer the correlated nature of an AF state. 
Optical spectroscopy is an invaluable experimental probe of correlated materials, providing key physical information 
such as the optical gap, the relative weight of low-energy Drude excitations in metallic systems, and quite importantly the transfers of spectral weights often observed in correlated materials when temperature or composition are  varied (for a recent review see Ref. \onlinecite{basov}). 
Spin degrees of freedom are however much less coupled to light than the charge so that 
the signatures of magnetism are expected to be comparatively weak. Zone-center magnons, for example, 
do not show up in the optical conductivity unless inversion symmetry is broken\cite{basov}. 
Furthermore, the key energy scale associated with magnetic ordering, the superexchange $J$ ($\sim D^2/U$ with $D$ being half  the bandwidth), is much smaller than the scale $U$ corresponding to the local matrix element of the screened Coulomb interaction, and hence generally significantly smaller than the optical gap itself ($\sim U-2D$). 

Despite these shortcomings, we show that clear signatures associated with antiferromagnetism are expected in the optical conductivity below the N\'eel temperature ($T_N$) when the system is in the strong-coupling (Mott insulating) regime. 
These signatures are twofold. 
First, there is a spectral weight transfer as one cools the system from above 
the N\'eel temperature down to low temperature. This transfer actually provides a diagnostics of the 
strong-coupling (superexchange) or weak-coupling (spin-density wave) nature of the antiferromagnetism. In the former case, 
spectral weight increase is observed, corresponding to a kinetic energy gain associated with the ordering. In the 
latter case, a spectral weight decrease takes place, corresponding to a kinetic energy loss (potential energy gain).  
For superconducting long-range order, similar diagnostics of weak vs. strong-coupling 
pairing have been discussed and probed experimentally in the context of cuprate superconductors\cite{superconducting} and 
for models with attractive interaction\cite{attractiveUale,attractiveU}.
Second, a multi-peak structure develops above the gap in the ordered phase at strong coupling. These peaks correspond to the    
spin-polaron excitations associated with the motion of a hole in the antiferromagnetic background. 
They have been previously discussed in the context of one-particle spectroscopy such 
as photoemission in Refs.\ [\onlinecite{volly},\onlinecite{sangiovanniPRB2006}]  and this work identifies their signature in the optical conductivity.

Finally, we consider the possibility of detecting these signatures experimentally, and 
discuss in detail the case of LaSrMnO$_4$ where we argue that some of these effects may already have been observed\cite{goesslingPRB77}. 

The outline of the paper is as follows: In Section \ref{Sec:model} we 
summarize the main aspects of our calculation of the optical conductivity for the Hubbard and $t$-$J$ model. Details on the DMFT calculations for the two models  are provided in Appendix  \ref{Sec:tJ} and  \ref{Sec:approxSE}, respectively.
In Section   \ref{Sec:results} we discuss the main theoretical results before we compare them to experiment in Section   \ref{Sec:comp}. A summary and conclusion is given in  Section   \ref{Sec:concl}.

%
%\begin{itemize}
%\item Not so much is expected to happen by crossing the N\'eel temperature in optics.
%\item The gap above and below $T_N$ will indeed be roughly given by $U$ and no dramatic changes are expected between the long-range ordered and the paramagnetic phases (after all, optics is not supposed to be the right probe for exploring magnetic properties).
%\item Here we show that instead there are very clear signatures of the crossing of $T_N$ if the antiferromagnetism is of strong-coupling nature. By means of Dynamical Mean Field Theory we show that a prototypical model for strongly correlated systems displays two features that are strikingly different above and below $T_N$: $i$) the emergence of a peak structure due to the formation of spin-polarons and $ii$) a pronounced temperature dependence of the spectral weight which is absent in the paramagnetic phase.  
%\item The frequency-dependence of the self-energy in DMFT allows one to detect within the same calculation the large energy scale associated to the gap, in turn to the Hubbard repulsion $U$, and a smaller one associated to a fine-spaced set of peaks, in turn to the antiferromagnetic super-exchange. 
%These two energy scales are indeed present in the optical conductivity of strongly-correlated antiferromagnetic compounds.
%\end{itemize}

\section{Models and methods}
\label{Sec:model}

In order to study the optical conductivity of strongly correlated antiferromagnets, we perform dynamical mean-field theory (DMFT) 
calculations for a half-filled Hubbard model, described by the following Hamiltonian:
\begin{equation}
H=-t\sum_{\langle ij\rangle \sigma }c_{i\sigma }^{\dagger }c_{j\sigma
}+U\sum_{i}n_{i\uparrow }n_{i\downarrow }  \label{H}.
\end{equation}
Here, $t$ denotes the hopping amplitude between nearest-neighbors, $U$ the
Coulomb interaction, and $c_{i\sigma }^{\dagger }$($c_{i\sigma }$) creates
(annihilates) an electron with spin $\sigma $ on site $i$; $n_{i\sigma
}\!=\!c_{i\sigma }^{\dagger }c_{i\sigma }$.
%AG
%I separated the general method from more specific details
% 
For simplicity, we use a semi-elliptical density of states of half-width $D$, corresponding to a Bethe lattice with infinite coordination. 
Our results however hardly depend on this specific choice. 
% 
%with a semi-elliptical density of the states (corresponding to a Bethe lattice in infinite dimensions with semi-bandwidth $D$). 
%
We vary the temperature $T$ from above to below the N\'eel temperature.
We therefore need the paramagnetic and the antiferromagnetically ordered solutions, which can both 
be calculated within DMFT\cite{DMFT,DMFTREV}.
 
It is known that the Hubbard model for large values of the local repulsion $U$ maps onto the $t\!-\!J$ model. 
For the specific model under consideration, the relation between $J$ and $U$ is the following: for an individual pair of sites we have $J_{\rm ind.}=4t^2/U$ and summed over all nearest neighbors $z$ we have $J = z J_{\rm ind.} = D^2/U$ (coined $J^*$ in Ref.\ \onlinecite{volly}).
The quantities of interest in the $t\!-\!J$ model in infinite dimensions can be expressed in the form of 
a simple continued-fraction\cite{volly,logan} while the Hubbard model requires a more computationally expensive solver. 
Therefore we will often use the $t\!-\!J$ solutions in the case of the antiferromagnetic phase. 
Indeed, the electron-removal part of the $k$-integrated spectral function of an half-filled Hubbard model below $T_N$ coincides up to higher order terms in $J$ with the spectral function of a single hole in the $t\!-\!J$ model.
The way we compute the optical conductivity for the $t\!-\!J$ model, and in particular how we get across the absence of an electron addition part of the spectrum of the  $t\!-\!J$ model, is explained in detail in the Appendix.

\begin{widetext}
Within DMFT the optical conductivity can be expressed exactly in terms of the convolution of two single-particle Green functions 
without vertex corrections \cite{DMFTREV,bluemer,khurana}. 
In the antiferromagnetic phase this reads \cite{zitzlerPruschke}:
%AG It seems to me there were typos in the equations, I made some changes, pls check.
\begin{eqnarray}
\label{optcond}\nonumber
\mathrm{Re}\,\sigma(\Omega)=\frac{e^2\hbar}{V\pi}\sum_\sigma \int d\omega \int_{-D}^D d\epsilon ~ \frac {f(\omega) - f(\omega + \Omega)}{\Omega} \frac{2}{\pi D^2} (D^2-\epsilon^2)^{3/2} \times \\
 \left[\mathrm{Im}G_\sigma^{AA}(\epsilon,\omega)\mathrm{Im}G_{\bar \sigma}^{AA}(\epsilon,\omega+\Omega)+ \mathrm{Im}G_\sigma^{AB}(\epsilon,\omega)\mathrm{Im}G_{\bar \sigma}^{AB}(\epsilon,\omega+\Omega)\right],
\end{eqnarray}
where $f$ denotes the Fermi distribution function, $G_\sigma^{AA}$ and $G_\sigma^{AB}$ are respectively the Green function of an electron with spin $\sigma$ hopping between sites of the same sublattice and between sites of the two different sublattices $A$ and $B$, in which the Bethe lattice is split in the AF phase.
\end{widetext}
%AGC I added a brief comment on the choice of the transport function. I dont know if there is a ref. to quote there 
% perhaps some paper by Millis et al ?
% We can quote the Ph.D. thesis of Bluemer and Tomczak, we did not find any better paper from Millis. 
In this expression, the quantity $\frac{2}{\pi D^2} (D^2-\epsilon^2)^{3/2}$ represents the product of the density of states of the Bethe lattice times the current vertex. This assumption insures that the f-sum rule associated 
with the optical conductivity holds, namely that its integral is proportional to the kinetic energy \cite{chung1998,chatto2000,bluemer2003,bluemer,tomczak}. 

%in the usual form

\begin{figure}[tb]
\includegraphics[width=8cm]{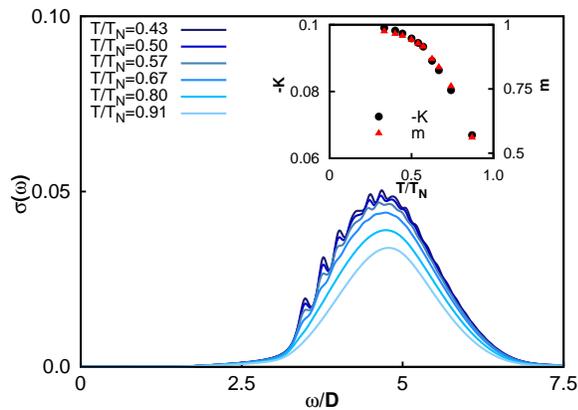}
\caption{(Color online) Temperature evolution of the optical conductivity  of the $t\!-\!J$ model at strong coupling ($U=5D$). Inset: temperature dependence of the absolute value of the kinetic energy ($-K$ in units of $D$) and of the staggered AF magnetization $m$.}
\label{Tevolution}
\end{figure}

\section{Main Results}  \label{Sec:results}
%AG Made some reshufflings below to better emphasize whats important. 
In Fig.~\ref{Tevolution} we show our main result, namely the temperature-dependence of the optical conductivity as one cools 
the system from close to the N\'eel temperature down to low temperature, in a regime of strong correlations, i.e., $U/D=5$. 
For temperatures slightly lower than $T_N$, the optical conductivity resembles that of the paramagnetic insulator: it has 
a gap of order $U-2D$, and the absorption peak above the gap is essentially featureless. 
% AG I commented out the sentence below which is not so important there. If you want to keep it, 
% displace it or make it a footnote.
%
%As we will describe below, the gap observed in the single-particle spectral function is smaller than that given by a static mean-field theory calculation, %namely it is reduced from $U$ to $U-2D$. As a consequence, the optical gap seen in Fig. \ref{Tevolution} is correspondingly given by $~U-2D$.
%
Upon cooling, two remarkable features characterize the optical signal: 
(i) the optical absorption strongly increases and (ii)  a multi-peak structure emerges. 
Such a structure becomes more and more visible as one goes down in temperature. 
Note that these changes occur without a visible variation in the position of the absorption edge (size of the gap). 
As discussed below, they should nevertheless be detectable by accurate experiments as a function of temperature. 

Later in this section we will explain the physical origin of the peaks in the optical conductivity by investigating 
the single particle spectral function and its evolution with $U$. At this stage, let us only stress that these peaks 
are by no means an artifact of the discretization required by our exact-diagonalization DMFT solver\cite{sangiovanniPRB2006}. 
To better compare with experiments, we used a Lorenztian broadening of $0.05D$ to plot the optical conductivity. This value is small enough to distinguish the first peaks but also large enough to mimic the smearing of the multi-peak structure due to experimental resolution.
% and to finite dimensionality effects beyond DMFT.

We now discuss in more details the two key effects we mentioned before. 

{\it i) Change of spectral weight through $T_N$.}
The increase of spectral weight upon cooling is characteristic of a Mott antiferromagnet at strong coupling.
Therefore it is the first gross feature one may look for in an optical experiment in order to establish whether the antiferromagnetism in the material is of strong-coupling nature, with preformed magnetic moments, or it is of weak-coupling nature, with a collective Fermi-surface instability.
%or not the antiferromagnetism in the material is of strong-coupling nature.
%
To understand this, we can use the fact that the integral of the spectral weight
%AG  
is directly proportional to the kinetic energy. % \cite{bluemer,tomczak,vdM}.
At strong coupling $U \gg D$, the ordered phase is stabilized by a gain in kinetic energy,  associated to the establishment of long-range coherence of the local magnetic moments.
In simple words the system gains kinetic energy by increasing the staggered magnetization: Coherent hopping processes take place in a staggered spin background are favored if the spin pattern is as close as possible to the N\'eel one \cite{sangiovanniPRB2006,attractiveU}. 
Thus the lower the temperature, the larger the staggered magnetization and in turn the larger the spectral weight.
%
%AG Quote the analogy also with U<0 and appropriate refs. 

%\begin{widetext}
\begin{figure*}
\includegraphics[width=12 cm]{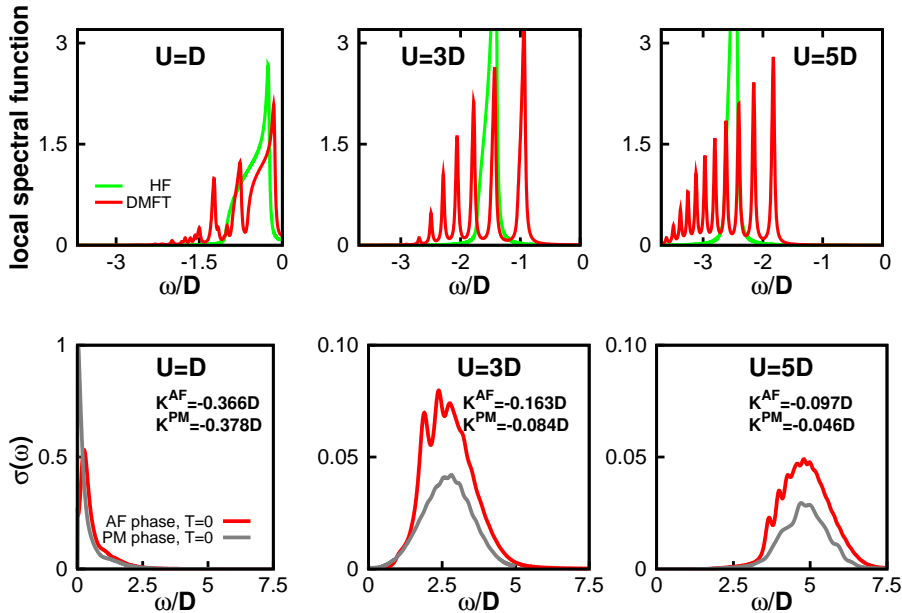}
\caption{(Color online) Lower panel:  Evolution of the optical conductivity of Hubbard model from weak (left) to strong coupling (right). Shown are both, AF phase and PM phase.  In the figure
the corresponding values of the optical integrals (=kinetic energy) are also reported, showing a change of the ``hierarchy'' between the kinetic energy scales of the two phases at intermediate coupling. \newline
 Upper panel: Corresponding spectral function with a  multi-peak (spin-polaron) structure. \cite{comm1} }
\label{Uevolution}
\end{figure*}
%We now discuss in more details these two key effects. 
%\end{widetext}

\begin{figure}[tb]
\includegraphics[width=8cm]{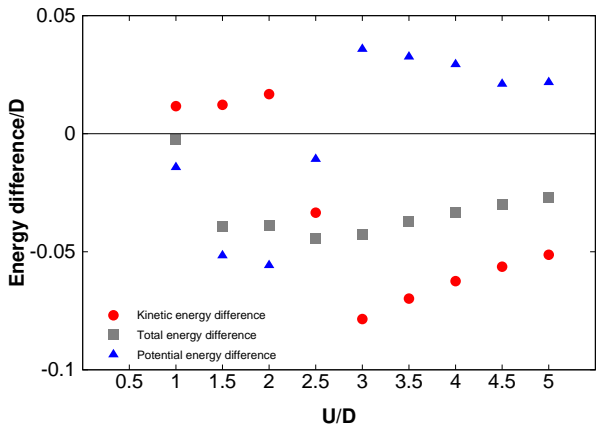}
\caption{(Color online) Difference of kinetic, potential and total energy between AF  and PM phase from weak to strong coupling. At strong coupling AF is stabilized by the kinetic energy and at weak coupling by the potential energy.}
\label{hierarchy}
\end{figure}

If we consider instead  the weak-coupling regime $U \ll D$, a totally different mechanism for the stabilization of the ordered phase takes place. 
The AF phase is stabilized by a (small) potential energy gain, as the onset of a non-zero order parameter is correctly described by the static mean field theory in this regime. These theoretical considerations are well confirmed by our numerical data shown in the bottom panel of Fig. \ref{Uevolution} where we compare $K^{AF}$ and $K^{PM}$, the kinetic energy of the ordered and disordered phase respectively.
The complete evolution of the kinetic and potential energy as a function of $U$ is shown in Fig. \ref{hierarchy}.
At strong coupling the antiferromagnetic order is clearly stabilized by the kinetic energy while at weak coupling ($U<2D$) the ordered phase is driven by the gain in potential energy. At intermediate values of the interaction strength, corresponding to the cross-over between weak and strong coupling,  we have a mixed regime in which both energy differences are negative \cite{attractiveUale,attractiveU}, i.e. where the onset of the AF is stabilized by both kinetic and potential energy.  
 
{\it ii) Spin-polaron peaks.}
To clarify the origin of the multi-peak structure, we compare in Fig. \ref{Uevolution}  the evolution from weak to strong coupling of the optical conductivity to that of the one-particle spectral function.
It is evident that the multi-peak structure of the spectral function\cite{volly,sangiovanniPRB2006} survives in the optical conductivity. 
The peculiar behavior of the spectral functions was first understood in a pioneering work by Strack and Vollhardt \cite{volly} for the $t\!-\!J$ model in infinite dimensions; and by some of us more recently within the Hubbard model in the antiferromagnetic phase by means of DMFT\cite{sangiovanniPRB2006}.
The multiple peaks arise from string-like excitations associated with a hole moving in a N\'eel background. In infinite dimensions the lifetime of such string excitations is infinite due to the absence of quantum spin-flip processes \cite{volly,metzner,sangiovanniPRB2006} which would act as `damage-repairing' processes, which means that the Heisenberg interaction reduces to the Ising one (see Appendix \ref{Sec:tJ} for further details).   This results in the formation of coherent but heavy quasiparticles, ``dressed'' by a spinon cloud which is commonly called a ``spin-polaron''.

The formation of the spin-polaron can be described only using methods (such as DMFT) that go beyond the static mean field of a spin-density wave 
in the mean-field Hartree-Fock approximation. This is illustrated in the upper panels of Fig.~\ref{Uevolution}, where we compare the evolution with $U$ of the dynamical (red line) and static mean-field (green line) results for the electron-removal spectrum. One clearly sees that the only feature of the static mean-field (Hartree-Fock) calculation is a Slater-like single peak centered around $-U/2$. The peak has weight $Z=1$ and its width is given by $J$, i.e. the width shrinks to 0 as $1/U$ upon increasing $U$. In the dynamical mean-field calculation, instead, the weight of the lowest lying excitation is much smaller ($Z$ is proportional to $J/D$) and the remaining weight is transferred to the higher-energy peaks. This makes a total bandwidth of order $2D$ rather then $J$, resulting in a smaller gap compared to the Hartree-Fock calculations.\cite{sangiovanniPRB2006}. 
% AG Displaced from above. 
This is why the gap in Fig.~\ref{Tevolution} is of order $U-2D$, while 
it would be exactly $U$ within static mean-field theory.

In the $t\!-\!J$ model in infinite dimensions (as well as in the Hubbard model within AF-DMFT, up to higher-order corrections in $t^2/U$) the position of the peaks can be analytically expressed through the zeros of the Airy functions \cite{volly} and the spacing between them goes as $J^{2/3}$. 
As we will also discuss in Section \ref{sec:concl} these analytic expressions hold strictly only in infinite dimensions.
Hence the applicability of our results to real materials, and in particular the emergence of spin-polaron peaks, strongly relies on the assumption that it is possible to neglect the quantum spin-flip processes. In principle this is not guaranteed for finite dimensional systems, therefore one may expect some of the peaks to broaden or even to be washed out, due to the coupling to {\em dispersive} spin waves \cite{sangiovanniPRB2006,spinpolaron}.
On the other hand, recent sophisticated calculations for the two-dimensional $t\!-\!J$ model give clear evidence for the survival of the first one or two spin-polaron peaks with a similar separation as in the infinite-dimensional calculation \cite{bonca}.
 Furthermore the case of a double layer antiferromagnet was recently considered in Ref. \onlinecite{zaanen}. There, it is shown that the inter-layer exciton made of one hole and one doublon in each layer displays confinement effects when the inter-layer coupling is smaller than the intra-layer one. In such a situation, the spin-polaron excitations discussed here may become particularly relevant. 

When considering three dimensional cases, where DMFT is typically more accurate  and finite-dimensionality effects are definitively weaker than in two dimensions\cite{digammaei},  spin-polaron features are expected to  be even more visible. In the following section, we discuss indeed the relevance of our results for the three dimensional manganite LaSrMnO$_4$.
      
%\begin{figure}[tb]
%\includegraphics[width=8cm]{./quinto.eps}
%\caption{(Color online) Behavior in temperature of the kinetic energy difference between AF and PM phase (in units of $D$). In the inset, for a comparison, it is shown the 
%behavior of the staggered magnetization in the AF phase. }
%\label{Fig:5}
%\end{figure}

\section{Comparison with experiments}
\label{Sec:comp}

\begin{figure}[tb]
\includegraphics[width=6cm,angle=-90]{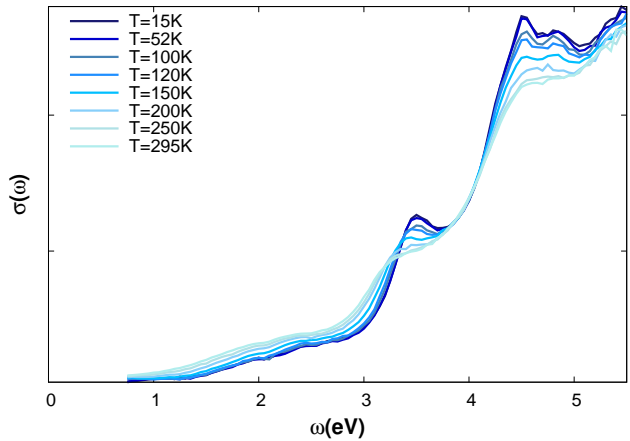}
\caption{(Color online) Optical conductivity of $\mathrm{LaSrMnO_4}$ between $0.75$ and $5.80$ eV for different temperatures. The raw data have been kindly provided by the authors of Ref. \onlinecite{goesslingPRB77}.}
\label{exp}
\end{figure}

On the basis of the previous discussion, the ideal material for the observation of the temperature evolution predicted above  would be a completely isotropic three-dimensional Mott antiferromagnet.
% with $G$-type spin pattern (i.e. antiferromagnetic in all directions) 
%and strong electronic correlations.
%
%Not many systematic $T$-dependent measurements have been performed hitherto on such a material.
% 
To our knowledge the most relevant set of measurements in this respect is that by G\"ossling, {\it et al.}, \cite{goesslingPRB77} on LaSrMnO$_4$, a three-dimensional $C$-type layered antiferromagnetic material with $T_N=133$K, while we are not aware of similar studies as a function of temperature on $G$-type antiferromagnets, which would be the closest realization of our large-coordination results.
In Fig. \ref{exp} we show the data of Ref. \onlinecite{goesslingPRB77} for the optical conductivity of LaSrMnO$_4$ along the $a$ axis at different temperatures. 
The two main structures around 3.5 eV and 4.5 eV have been explained in terms of  multiplet split transitions from a $d^4$ to a $d^5$ configuration of the Mn.
Our main interest instead is focused on the fine structure arising upon cooling on top of both of these main structures.
%The authors of Ref. \onlinecite{goesslingPRB77} mentioned that a detailed analysis of the spin-spin correlation function would be required in order to understand the observed temperature behavior, but otherwise did not provide a precise explanation. 
%

In the light of our theoretical analysis, we can directly relate this observation with the physics of the spin-polarons described in the previous section.
%
%AG Changed a bit the phrasing below. 
Since the two structures are well separated, we  shall make a simplifying assumption (not meant to be quantitative, but sufficient to reveal the main qualitative aspect of the phenomenon). Namely, we shall assume that we can adopt for each $d^4 \rightarrow d^5$ transition a simple single-band Hubbard model description.

Since the separation between the multiplet peaks is larger than the energy scale associated with the spin-polaron excitations, we expect to be able to detect the above-mentioned increase of spectral weight around each of the main absorption peaks and the consequent appearance of spin-polaron structures when going below the N\'eel temperature.

All curves below $T_N$ display, in the same region, a very similar low-frequency tail (below $\sim3$~eV) while in that region the higher-temperature measurements show a gradual increase of spectral weight that we attribute to a more standard thermal broadening effect. It is likely that such thermal effects are visible also in the low-energy multiplet peak at 3.5 eV, which lies closer to this tail. This makes this spectral feature less visible with a first distinguishable excitation shifting with temperature unlike what happens for the second multiplet structure at 4.5 eV.
\begin{figure}[tb]
\includegraphics[width=6cm,angle=-90]{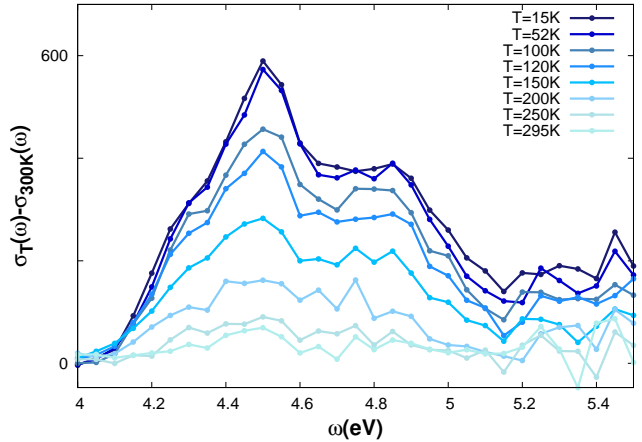}
\caption{(Color online) Difference between the optical conductivity of $\mathrm{LaSrMnO_4}$ between $4.0$ and $5.5$ eV at different temperatures with respect to the optical conductivity at $T=300$K. Data by G\"ossling, {\it et al.}, \cite{goesslingPRB77}.}
\label{exp_zoom}
\end{figure}

We focus therefore on the high-energy spectral structure around 4.5 eV, which seems to be unaffected by the temperature broadening of the low-frequency data, and we zoom the plot to focus on this feature.
%Even though all curves are very similar between $~3.8$ and $~4.2$eV, i.e. the tail for the $4.5$eV peak is hardly temperature dependent, we cannot exclude further temperature effects beside the AF ones we describe here also in the region of frequency relevant for the $4.5$eV peak.
In order to highlight the temperature effect, we display in Fig. \ref{exp_zoom} the difference between the optical conductivity at a given temperature and  the conductivity at the highest measured temperature (300K).

\begin{figure}[tb]
\includegraphics[width=6cm,angle=-90]{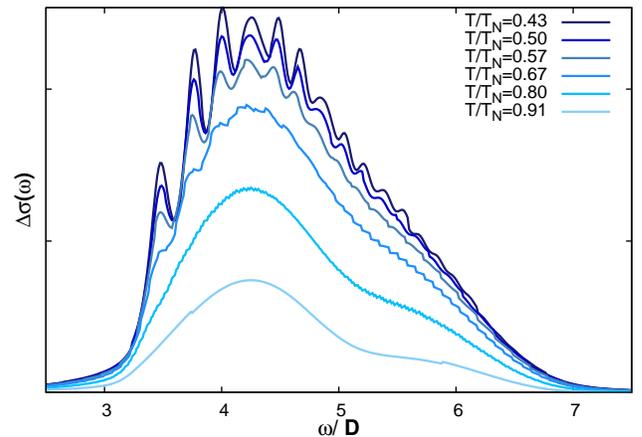}
\caption{(Color online) Difference between the computed optical conductivity at different temperatures and the one at $T_N$,\cite{comm2} showing a very similar structure as the experiment in Fig.\  \ref{exp_zoom}.}
\label{theo_zoom}
\end{figure}

Fig. \ref{exp_zoom} and its theoretical counterpart (Fig. \ref{theo_zoom}) endorse our physical interpretation of the peaks in the optical signal. 
Although the frequency grid in the experimental data is 0.05 eV, at least two well-spaced peak-structures emerge below $T_N$ and, as expected within our spin-polaron picture, the lower the temperature the weaker the temperature dependence becomes. 
Indeed between 52 and 15K there is hardly any difference in the optical conductivity in this frequency region, as one can expect from the behavior of the staggered magnetization (see inset to Fig. \ref{Tevolution}), which is almost saturated far below $T_N$.
The appearance of the spin-polaron fine structures become clear below $T_N=133$K. Yet, the polaronic features start to be visible already in the 150K curve. We attribute this to fluctuation effects above $T_N$: even if long-range order is not yet fully developed, a sufficiently long coherence length of the spin-spin correlations is enough to sustain a non-fully developed spin-polaron mechanism.

Similar features as the ones we have described are present in the optical absorption of another material: KCuF$_3$. For temperatures lower than the magnetic ordering temperature of this compound a multi-peak structure is indeed observed (see Fig. 2b of Ref. \onlinecite{deisen}), for which an interpretation in terms of spin-waves has been given. 

We have therefore found in nature what we expect from our model calculation: A clear signature of antiferromagnetism arising from strong correlations in the optical conductivity of a strongly correlated material which emerges as the system is cooled below $T_N$. 
If the material is sufficiently three-dimensional and if the resolution of the experiment is good enough, a clear signal with a multi-peak structure can be observed.
Another way of revealing the same effect would be to compare the increase of spectral weight to the measured staggered magnetization as a function of $T$.

%In addition, dimensionality effects are also expected to smear out them (see discussion in Section \ref{sec:concl}). 

\section{conclusions} \label{sec:concl}\label{Sec:concl}

We have calculated the optical conductivity of a strongly correlated antiferromagnet described by the Hubbard or the $t\!-\!J$ model. Within dynamical mean-field theory, we have shown that the optical conductivity is characterized by the presence of equally spaced spin-polaron peaks which appear right below the N\'eel temperature and become more and more pronounced as the temperature is reduced.
The onset of antiferromagnetic ordering is stabilized by a kinetic energy gain in strong coupling superexchange magnets, while at weak-coupling antiferromagnetism is due to Fermi-surface nesting and it is stabilized by a static mean-field like potential energy gain. 

In the strong coupling regime, the sharp peaks associated to spin-polarons originate from string-like excitation created by the motion of a hole in an antiferromagnetic background, and we can safely distinguish them from artificially sharp peaks characteristic of exact diagonalization solutions of DMFT.

Even if our formal derivation of the spin polaron peak relies on the infinite-coordination limit in which DMFT becomes exact, we expect at least part of these features to survive in finite dimensions, and in particular in the three dimensional case, in which DMFT typically proves accurate \cite{digammaei}. We notice that at least the lowest-energy excitation peak is found also in two dimensions, as shown by different approaches\cite{bonca}, testifying that the physics behind our DMFT results survives when the large dimensionality limit is far.

If we look for experimental realizations of this scenario, we should take into account temperature effects and the finite experimental resolution, which can further smear out some of the fine structures. In this light, it is remarkable that an experimental study on LaMnSrO$_4$\cite{goesslingPRB77} reveals a multi-peak structure whose form and temperature behavior are in very good agreement with our calculations on top of the multiplet structure.

Moreover, considering the recent work by Rademaker {\sl et al.} \cite{zaanen}, optical spectroscopy performed  along the $c$-axis in layered systems may be a promising tool to detect and investigate the occurrence of the spin-polaron physics.

An even stronger signatures of the spin-polaron effects are to be expected in three dimensional correlated materials which present a three-dimensional $G$-type antiferromagnetism, like SrMnO$_3$, CaMnO$_3$, BaMnO$_3$, calling for future IR-spectroscopy investigations of the PM-AF transitions in these compounds.

\textit{Acknowledgments.} We thank Philipp Hansmann, Maurits Haverkort,  Dirk van der Marel and Andrei Pimenov for fruitful discussions. We are also indebted with Erik Koch, Olle Gunnarsson and Claudio Castellani for their contribution to the early stage of the AF-DMFT coding and interpretation. C.T. acknowledges financial support from Research Unit FOR 1346, project ID  I597-N16 of  the Austrian Science Fund (FWF). M.C. was supported  by European Research Council under FP7/ERC Starting Independent Research Grant ``SUPERBAD" (Grant Agreement n. 240524);
 AT through the FWF project I610-N16 and KH through the SFB ViCoM  FWF F4103-N13. Calculations have been done on the Vienna Scientific Computer (VSC).

\section{Appendix}

In this Appendix we give all details about how to construct an approximated expression for the DMFT self-energy of the half-filled Hubbard model on the basis of the DMFT solution for the corresponding $t\!-\!J$ model both in the antiferromagnetic and the paramagnetic phase. We will discuss also the validity of this procedure by exploiting the mapping of the (half-filled) Hubbard Hamiltonian onto the $t-J$ one\cite{mappingHubbardtJ} for $U \gg D$ and by   comparing our results for the self-energy and the optical spectra extracted from the DMFT solution of the $t-J$ model with the correspondent ones for the Hubbard model. The latter are available with high enough numerical precision.

\subsection{DMFT for the $t\!-\!J$ model}
\label{Sec:tJ}

The main motivation to use the $t\!-\!J$ results for approximating the DMFT self-energy  stems, as mentioned in Sec. II, from the simplicity of the algorithm for solving the DMFT problem of one hole in the $t\!-\!J$ model with (complete or partial) AF-N\'eel order\cite{volly} as opposed to the bigger numerical effort needed for the Hubbard model. %In fact, as is was discussed in Ref. \onlinecite{volly}, taking the limit of infinite dimensionality (i.e., DMFT) is way easier (and more precise) for the $t\!-\!J$ model than for the Hubbard model. 
Specifically for  one-hole at $T=0$ one finds (for more details see Ref.\ \onlinecite{volly}) that for $d \rightarrow \infty$

\begin{enumerate}

\item the quantum spin-fluctuations associated to the term proportional to $S_{i}^xS_{j}^x+ S_{i}^yS_{j}^y$ of the SU(2)-symmetric Heisenberg interaction are completely frozen. In other words, the Heisenberg coupling of the $t\!-\!J$ Hamiltonian is reduced to an Ising coupling, whose ground state is the AF-N\'eel state with a perfectly staggered spin configuration, i.e, with full staggered magnetization ($m=\frac{1}{N}\sum_{r_i} (n_{i\uparrow} -n_{i,\downarrow})(-1)^{r_i}=1$, where $N$ represents the total number of sites);

\item the ``retraceable path'' approximation becomes rigorously exact: In a perfect AF-N\'eel state, the local physics (e.g., the local Green function $G(\omega)$) is built only by hopping processes in which the one hole comes back to the starting site following  precisely the same path. In fact, any non-retraceable path is associated with an higher-energy cost, as the hole movement would result in a path of misaligned spins in the AF-N\'eel background. This implies that the geometry of the underlying lattice becomes irrelevant in $d\rightarrow \infty$: If only retraceable paths matter, every geometry becomes topologically equivalent to the Bethe lattice, a tree-structure which by construction only allows for retraceable paths.
 
\end{enumerate}
As a consequence at $T=0$ the discrete energy levels characterizing the analysis of one-hole dynamics in the AF-N\'eel background of the $t\!-\!J$ model can be related to the simple quantum mechanical problem of one particle in a step potential $V(n)=n J$, being $n$ the number of misaligned spin created by the hopping processes of the hole. This allows in turn to obtain a compact continued-fraction expression for the retarded Green function $G_{t\!-\!J}(\omega)=G_{t\!-\!J}(\omega +i0^+)$ (see Ref. \onlinecite{volly})
\begin{eqnarray}
G_{t\!-\!J}(\omega) &=& \left[\omega - \frac{D^2}{4}  G_{t\!-\!J}(\omega - \frac{1}{2}J) \right]^{-1} =
\label{eqapp:volly} \\
 &=& \frac{1}{\omega - \frac{D^2}{4}\frac{1}{\omega - \frac{J}{2} - \frac{D^2}{4} \frac{1}{\omega - J - \frac{D^2}{4}\frac{1}{\cdots}  }}}
\end{eqnarray}
which can be solved with high numerical accuracy by simply truncating the continued fraction to the desired level of precision, and then building up the full expression iteratively. Complications arise at finite $T$: In particular point 2 is in principle no longer valid, as thermal fluctuations will reduce the value of the staggered magnetization, allowing for a contribution of non-retraceable paths to the local physics. However, we can easily circumvent this problem, if we consider explicitly the Bethe lattice case for our DMFT calculations. This is, on the one hand, a reasonable choice, as the finite bandwidth of the Bethe lattice reflects more directly the realistic situation of finite bandwidths in real materials, and, on the other hand, it preserves by construction the validity of the ``retraceable path'' condition at all temperatures.         
Working with the Bethe lattice allows, in turn, for a straightforward generalization of the continued-fraction expression for $G(\omega)$ to the finite temperature case, along the lines of the exact derivation of Ref.\onlinecite{logan}. Specifically, one can easily generalize Eq. (\ref{eqapp:volly}) to finite temperatures by taking into account the probability $P$ (or $1-P$) of finding correctly aligned (or misaligned) spins along the retraceable path of the hole. The probability $P$ as a function of $T$ is easily expressed as $P= \frac{1}{2}(1+ m(T))$ in terms of the staggered magnetization $m(T)$. The latter, because of point 1 above, can be directly obtained by the simple mean field Curie-Weiss self-consistent equation $m(T)=\mbox{tanh}\left(\frac{T_N}{T} m\right)$, with $T_N=\frac{1}{4} J$. The value of $P$, when smaller than $1$, will affect the self-consistent expression for the retarded Green function:
\begin{eqnarray}
G_{t\!-\!J}(\omega)  & =  &  \left[\omega + \! \frac{D^2}{4}\left(P G_{t\!-\!J}(\omega \! - \! \omega_P)  \right. \right. \nonumber \\ & + & \left. \left. (1\! - \!P) G_{t\!-\!J}(\omega + \omega_P)\right) \right]^{-1}  
\label{eqapp:logan}
\end{eqnarray}
being $\omega_P= \omega_P(T)= \frac{1}{2}m(T)J$, and $\omega = \omega +i0^+$. Note that this expression is exactly reduced to Eq.\ (\ref{eqapp:volly}) in the limit $T \rightarrow 0$, as $m(T=0)=P(T=0)=1$.
The  additional terms in Eq.\ (\ref{eqapp:logan}) at finite $T$ result in a increasing   proliferation of new (smaller) peaks no longer exactly matching with the original multi-peak structure of the continued-fraction expression at $T=0$. As a consequence, by increasing $T$, one observes a gradual smearing out of the $\delta$-like peak structure of Eq.\ (\ref{eqapp:volly}), which is completely washed out at $T \geq T_N$, when one has $m=0, P=\frac{1}{2}$ and  the spectral representation $G(\omega)$ is reduced to the characteristic semicircular shape of the non-interacting Bethe-lattice case. This result can be easily understood, because the Heisenberg interaction term, which is {\sl non-local} in the site indices $i$,$j$, becomes irrelevant due to the $d\rightarrow \infty$ scaling in the paramagnetic phase (i.e., for $m=0$) and, therefore, for $T \geq T_N$ one gets the same spectral function of the case $J=0$, independently from the actual value of $J$.      

\subsection{Approximation for the DMFT self-energy of the Hubbard model}
\label{Sec:approxSE}

After having recalled how to solve exactly the DMFT problem of one hole in the $t\!-\!J$ model, we will discuss here how to approximate the DMFT self-energy for the Hubbard model in the limit of intermediate-to-strong coupling by exploiting this solution. In fact, it is known \cite{mappingHubbardtJ} that in the limit $U \gg D$ the Hubbard model can be mapped onto a corresponding $t\!-\!J$ model with $J=\frac{D^2}{U}$. When using the $t\!-\!J$ results to approximate the Hubbard DMFT self-energy or the ${\bf k}-$integrated spectral and Green functions in the limit of $U \gg D$, one should not forget that the exact ``mapping'' between the two models is obtained  by projecting out the doubly occupied sites. This evidently means that the $t\!-\!J$ model cannot describe the whole frequency range of the Hubbard model, even in the limit of $U \gg D$, but, at most, only ``half'' of it. This becomes quite obvious, when explicitly considering the  ${\bf k}-$integrated spectral functions: The gap between the Hubbard bands is originated by the energy difference ($\sim U$) between single occupied and double occupied/empty sites, but this energy difference is exactly what is projected out in the mapping onto the $t\!-\! J$. Therefore, when computing the DMFT Green function for one {\it hole} in the $t\!-\!J$ model, the associated spectral function will be related to electron-removal ($\omega < 0$, i.e. photoemission) part  of the ${\it k}-$integrated spectral function of the  Hubbard model: More precisely, taking into account the above mentioned energy shift of order $U$ (not explicitly included in the $t\!-\!J$ Hamiltonian), it will correspond to the {\it lower} Hubbard band. Similarly, if one considers the problem of one {\it electron} added to the $t\!-\!J$ model, the associated spectral function would correspond to the  {\it upper} Hubbard band of the Hubbard model spectrum.

\begin{figure}[t!]
\includegraphics[width=5cm,angle=270]{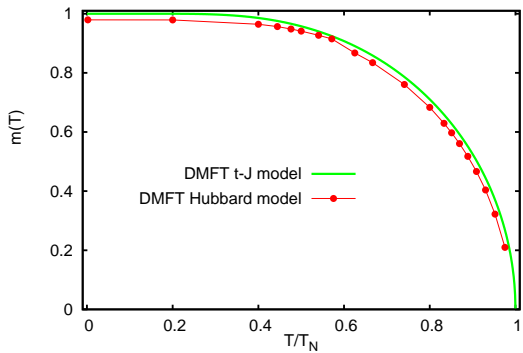}
\caption{(Color online) Comparison between the staggered magnetizations $m$ as a function of temperature obtained by means DMFT calculations for the  Hubbard model (with $U=5D$) and for the corresponding $t\!-\!J$ model: Due to the residual (but still finite) presence of double occupied/empty sites in the AF-ground state of the Hubbard model, the magnetization (as well as the value of $T_N$) is slightly lower in the former case, and -at difference with the $t\!-J\!$ case- it never reaches $1$ (complete magnetization), even for $T \rightarrow 0$.}
\label{figapp:magn}
\end{figure}

These considerations are easily translated in a simple practical implementation: In fact, considering the symmetry with respect to the change $\omega \leftrightarrow -\omega$ for both the paramagnetic and the antiferromagnetic case, one can obtain the $t\!-\!J$-like approximation of the ${\bf k}-$integrated  retarded Hubbard Green function $G_{\sigma}(\omega + i0^+)$ on a given sublattice (for the sake of definiteness we identify with $\sigma = \uparrow$ the majority spin component on the AF order on the chosen sublattice) directly from the retarded Green function (or the spectrum) $G_{t\!-\!J}(\omega)$ ($A_{t\!-\!J}(\omega)$) of the one-hole $t\!-\!J$ problem as
\begin{eqnarray}
G_{\uparrow}(\omega) \! &  = & P G_{t\!-\!J}(\omega \! + \!\frac{U}{2}) + (1\!-\!P) G_{t\!-\!J}(-\omega \! +\! \frac{U}{2}) \label{eqapp:mixspectra1}\\
G_{\downarrow}(\omega) \! & = & \! (1\! -\! P) G_{t\!-\!J}(\omega \! + \!\frac{U}{2}) + P G_{t\!-\!J}(-\omega \! +\! \frac{U}{2}) 
\label{eqapp:mixspectra}
\end{eqnarray}
The essential ingredients of Eqs.\ (\ref{eqapp:mixspectra1}) and (\ref{eqapp:mixspectra}) are, hence, two: (i) the spectral (Green) function of the one-hole problem of the $t\!-\!J$ model, and (ii) the  probability $P$  to find (on the chosen site) the spin correctly aligned with the AF-N\'eel background. A word of caution is due at this point: As the value of $P$ is defined in terms of the staggered magnetization $m(T)$, the latter naturally represents the most appropriate ``link'' between the DMFT $t\!-\!J$ and Hubbard Green functions. In fact, in using the $t\!-\!J$ approximation for the DMFT spectra of Hubbard model, one should keep in mind that the mapping is rigorously exact only in the limit of $U \rightarrow \infty$, otherwise, corrections $\sim \frac{D^3}{U^2}$ have to be expected: From a more physical perspective, the presence of doubly occupied/empty sites in the ground state of the Hubbard model stays indeed {\sl finite} (though becoming smaller and smaller for increasing $U$) at any value of the Hubbard interaction. As a consequence, for any given set of parameters $U$ and $T$, the staggered magnetization of the Hubbard model will be slightly {\sl lower} than that for the $t\!-\!J$ model, as it is also found in our DMFT data, plotted in Fig. \ref{figapp:magn}.

It should be clear, hence, that the best agreement between the  Hubbard  model (for a given $U$) and the  corresponding $t\!-\!J$ (with $J=\frac{D^2}{U}$)  approximation is found for sets of Hubbard and $t\!-\!J$ Green/spectral function with the {\sl same} magnetization values (which are achieved, according to Fig. \ref{figapp:magn}, for slightly different temperatures\cite{noteTmismatch}).  

\begin{figure}[t!]
\includegraphics[width=8cm]{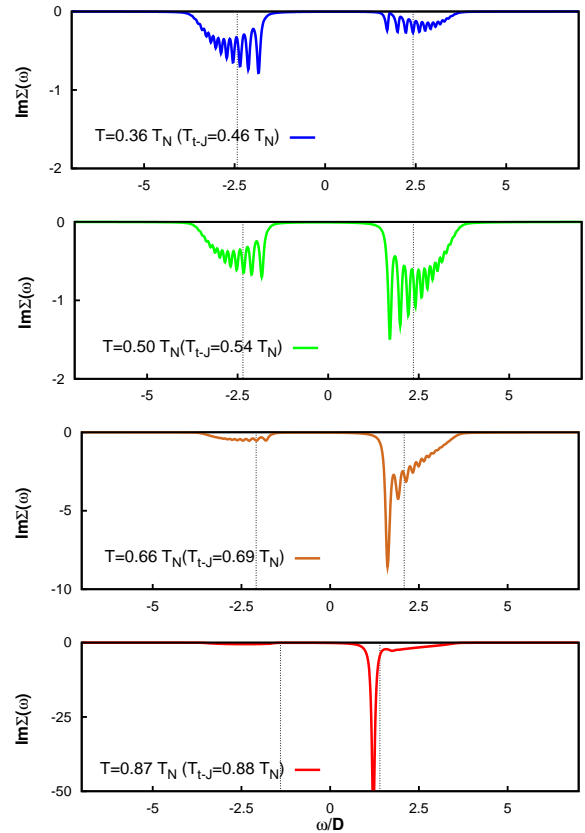}
\caption{(Color online) Temperature evolution of the DMFT self-energy of the Hubbard model, calculated from the DMFT Green function of the $t\!-\!J$ model, according to Eqs.\ (\ref{eqapp:mixspectra}) and \ref{eqapp:self} for the specific case $U=5D (J=0.2 D)$. Shown is the spin-up self energy on the sites with spin-up majority; the spin-down self energy is the particle-hole counterpart (obtained by mirroring $\omega\rightarrow -\omega$).
 As for the requirement of an equal staggered magnetization between the Hubbard and the $t-J$ model, the corresponding temperature of the $t\!-\!J$ case are reported in the legend as $T_{t-J}$ . The dotted vertical lines are located at $\omega=-\frac{U}{2}m$ and $\omega=+\frac{U}{2}m$ i.e., where in a static mean-field context one would find the two peaks of the self energy.}
\label{figapp:self}
\end{figure}

Having defined the equations (and the conditions) for best approximating the Hubbard DMFT Green functions with the $t\!-\!J$ model ones, it is straightforward to extract from these the correspondent DMFT expressions for the (spin-dependent) Hubbard self-energies $\Sigma_{\sigma}$, via the DMFT self-consistency equations (generally written to include the AF order\cite{DMFTREV}):
\begin{eqnarray}
\Sigma_{\uparrow}(\omega)  & = & \omega +\mu - \frac{D^2}{4}[G_{\downarrow}(\omega)]^{-1} \\
\Sigma_{\downarrow}(\omega)  & = & \omega +\mu - \frac{D^2}{4}[G_{\uparrow}(\omega)]^{-1},
\label{eqapp:self}
\end{eqnarray}   
where the chemical potential $\mu$ is set fixed to $\frac{U}{2}$ (particle-hole symmetric case).
Representative results of our DMFT $t\!-\!J$ approximation for the imaginary part of our  Hubbard self-energy ${\rm Im} \Sigma_{\uparrow}(\omega)$, computed according to Eqs.\ (\ref{eqapp:mixspectra1}) - (\ref{eqapp:self}) are reported in Fig.\ \ref{figapp:self}. The main feature of the $T-$dependence of $\mathrm{Im} \Sigma_\uparrow(\omega)$ well agree with the corresponding DMFT data for the Hubbard model shown in Ref. \onlinecite{sangiovanniPRB2006}, supporting the validity of the $t\!-\!J$ approximation. In particular, if $U \gg D$, both in Hubbard, as well as in its  $t\!-\!J$ approximation, one finds that for $T \ll T_N$  (upper panel of Fig. \ref{figapp:self}) $\mathrm{Im} \Sigma_\uparrow(\omega)$ is characterized by a multi-peak structure for $\omega < 0$ which reflects that of the spin-up spectral function, and, hence, the above-mentioned spin-polaron physics.
 Remarkably the ``magnitude'' of  $\mathrm{Im} \Sigma(\omega)$, e.g. if quantified by the value of its frequency integral, is not particularly large for low-$T$, implying a certain degree of ``coherence'' of the spin polaron excitations in this regime\cite{noteQPAF}. Note that, except for the case of full polarization a much weaker multi-peak structure is also visible for $\omega > 0$, roughly speaking where the upper Hubbard  would be located in the PM phase. 

By increasing $T$, one observes a rapid strengthening of such ``secondary'' multi-peak structure at  $\omega > 0$, which becomes quickly predominantly (second and third panel of Fig. \ref{figapp:self}). More specifically, one notes that the lower energy peak of the ``secondary'' ($\omega > 0$) structure of $\mathrm{Im} \Sigma(\omega)$, is strongly enhanced with increasing $T$,
 and at the same time undergoes a constant ``softening'', as its position (inside the spectral gap) tends to go to zero frequency for $T \rightarrow T_N$ (lower panel of Fig.  \ref{figapp:self}). In the limit of $T= T_N$, when $\Sigma_\uparrow(\omega) = \Sigma_\downarrow(\omega)$, this (now extremely strong peak) would be located exactly in the center (that is at $\omega=0$), ensuring the existence of the Mott-Hubbard gap $\sim U -2D$ also in the absence of any long range magnetic order. This can be compared with the exact solution of the atomic limit ($D=0$), for which one can easily compute the Green function\cite{DMFTREV}, e.g. for the spin $\uparrow$ sector
\begin{equation}
G_\uparrow(\omega)= \frac{\frac{1}{2}(1-m)}{\omega + i0^+ + \frac{U}{2}} + \frac{\frac{1}{2}(1+ m)}{\omega +i0^+ -\frac{U}{2}}
\label{eqapp:Gat}
\end{equation}
and, from this (and the corresponding $G_\downarrow(\omega)=G_\uparrow(-\omega)$), calculate the self-energy
\begin{equation}
\Sigma_\uparrow(\omega)= \frac{U}{2}\left[\frac{\frac{U}{2}-m\omega}{\omega+i0^+ -m\frac{U}{2}}\right]
\label{eqapp:Gat2}
\end{equation}
 which evidently displays a peak in its imaginary part located at $\omega_{peak} = m \frac{U}{2}$ (i.e, a ``softening'' for increasing $T$/decreasing $m$) of weight $w_{peak}= \pi \frac{U^2}{4}[1-m^2]$ (i.e., of increasing strength for increasing $T$/decreasing $m$). At $T=T_N$, one gets eventually $\Sigma_\uparrow(\omega) = \Sigma_\downarrow(\omega) = \frac{U^2}{4(\omega + i0^+)}$, whose divergence in $\omega=0$ directly corresponds to the spectral gap of (exact) size $U$. In comparison with the atomic limit, however, beyond this huge central peak in Im$\Sigma(\omega)$, one observes the formation of two almost semicircular structures at $\omega = \pm \frac{U}{2}$ of width $\sim D$, which can be interpreted as an hallmark of the incoherent nature of the electronic excitation of the Hubbard band in the paramagnetic phase.      
 \subsection{Accuracy of the approximation}

After having discussed the results obtained by approximating the Hubbard DMFT self-energy starting from the $t\!-\!J$ one, and having noted that for $U \gg  D$ its structure is coincident with that of the exact Hubbard model within small corrections (roughly of order $\frac{D^3}{U^2}$), here we will explicitly compare DMFT data for the optical conductivity in order to show the accuracy of our $t\!-\!J$ approximation for the quantity we are more interested in this paper.

\begin{figure}[tb]
\includegraphics[width=8cm]{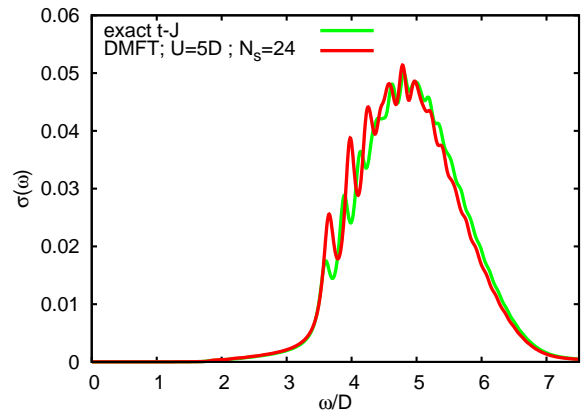}
\caption{(Color online) Comparison between the DMFT optical conductivity of Hubbard model (impurity solver: Lanczos with $N_s=25$) and corresponding $t\!-\!J$ model. }
\label{Hubb-tJ}
\end{figure}

We report in Fig.\ \ref{Hubb-tJ} our numerical data for the optical conductivity of the AF-phase, namely for $U=5D$ at $T=0$ calculated inserting the DMFT Hubbard self-energy and the correspondent $t\!-\!J$ approximated one in Eq.\ (\ref{optcond}). 
At $T=0$, in fact, the possibility to use the Lanczos as an impurity solver, allows for a better resolution of the multiple peak structure of the Hubbard self-energy and, therefore, for a better test case to evaluate the accuracy of the $t\!-\!J$ approximation for the optical conductivity of the Hubbard model.

In fact, while Lanczos already allows a larger number of bath sites than the full exact diagonalization, in the AF case for $T \ll T_N$ the computational cost of increasing the number of bath sites can be further reduced. 
Specifically it is reasonable to assume that, in the AF ground state, also the bath electrons will be almost fully polarized. 

Bearing in mind that the total number of electrons with spin up ($N_\uparrow$) and with spin down ($N_\downarrow$) are 'good' quantum numbers, the Hamiltonian can be decomposed in smaller blocks, each associated with a \emph{spin sector}, whose size is given by: 
%The dimension of the Hilbert space associated with $N_\uparrow$ spins up and $N_\downarrow$ spins down (and therefore the size of the matrices to diagonalize in each spin sector) in the half filled case is:
\begin{eqnarray}
\mathrm{dim}(\mathcal{H}_{N_\uparrow,N_\downarrow})&=&\mathrm{dim}(\mathcal{H}_{N_\uparrow})\times\mathrm{dim}(\mathcal{H}_{N_\downarrow})\\
&=&\binom{N_s}{N_\uparrow}\times\binom{N_s}{N_s-N_\uparrow},
\label{block_size}
\end{eqnarray}
where $N_s$ represents the total number of electron levels, and $\mathrm{dim}(\mathcal{H}_{N_\sigma})$ the dimension of the Hilbert space associated with $N_\sigma$ electrons of spin $\sigma$. 
Since the bath is strongly polarized, in the Lanczos algorithm we only need to take into account those spin sectors in which the big majority of the fermions of the bath has either spin up or spin down. That means that we only need to diagonalize the smallest blocks composing the Hamiltonian. 

This is very different from the PM phase, where we need to take into account mainly those sectors where the number of spin up fermions and spin down fermions is comparable, and hence the size of the matrices to diagonalize is much bigger. 
In practice, exploiting the property of a fully polarized bath, in the AF phase we can reach with a reasonable computational effort a number of electrons as high as $N_s=25$, while in the PM case we could reach only $N_s=14$

Going back to the analysis of Fig. \ref{figapp:magn}, our data show that, at $U = 5D$, the overall agreement between the peaky structure of the optical conductivity of the $t\!-\!J$ approximation and that of the Hubbard model is rather good, as both the peak positions and intensity are well captured also in the $t\!-\! J$ scheme. 
This result justifies, for these values of $U$, the usage of the DMFT solution of the one-hole problem in the $t \!- \!J$ model as an accurate approximation, e.g., also when considering the finite $T$ regime. In fact, when increasing $T$, Lanczos or full ED solution of DMFT are no longer accurate enough for the computation of multi-peak spectral functions with sufficient resolution.

%\vspace{25 cm}


\begin{thebibliography}{99}
\bibitem{comm1} Note, except for the peaks around $\omega=-1.5D$ to $-1D$ in the upmost left panel, all peaks are not numerical artifact but, as described in the text, have a precise physical correspondence to spin-polaron peaks.
\bibitem{comm2}
 Note  within the $t\!-\!J$ model the temperature dependence of the optical conductivity above $T_N$ is just the one coming from the Fermi functions in Eq.\ (\ref{optcond}). Within the Hubbard model we would instead get some temperature dependence, however only appreciable on a rather big energy scale.
\bibitem{noteTmismatch} For instance, at $U=5D$, the maximal magnetization (at $T=0$) for the Hubbard model is $m=0.979$, which correspond to the $t-J$ solution at $\beta D= 46.6$.  This slight difference in temperature is naturally reduced with increasing $U$.
\bibitem{noteQPAF} In other words, this means that if the chemical potential were rigidly shifted to the value of the first (low-energy) AF magnetic peak, this would behave as a coherent quasiparticle excitation  (though strongly renormalized by $Z \sim \frac{D^2}{U}$). A similar situation would occur, ``mutatis mutandis'', in the case such an excitation were activated by an external perturbation of frequency $\omega$ of the size of the spectral gap, e.g., by an electromagnetic field, when measuring the optical conductivity $\sigma(\omega)$.
\end{thebibliography}
\end{document}